\documentclass{aa}
\usepackage{amssymb}
\usepackage{graphicx}
\usepackage{hyperref}

\def\asec{\ifmmode ^{\prime\prime}\else$^{\prime\prime}$\fi}

\def\degs{\ifmmode ^{\circ}\else$^{\circ}$\fi}
\def\amin{\ifmmode ^{\prime}\else$^{\prime}$\fi}
\def\asec{\ifmmode ^{\prime\prime}\else$^{\prime\prime}$\fi}
\def\fm{\hbox{$.\!\!^{\rm m}$}}            
\def\farcs{\hbox{$.\!\!^{\prime\prime}$}}  

\def\psr{PSR~B0950+08}
\def\degs{\ifmmode ^{\circ}\else$^{\circ}$\fi}
\def\amin{\ifmmode ^{\prime}\else$^{\prime}$\fi}

\def\hst{{\sl HST\/}}
\def\rosat{{\sl ROSAT\/}}
\def\asca{{\sl ASCA\/}}

\def\widerul{\vrule height 2.5ex width 0ex depth 0ex}

\def\twolines#1#2{
\renewcommand{\arraystretch}{0.8}
\begin{tabular}{@{}l@{}}
#1 \vrule height3.2ex width0ex \\ #2 \\[.7ex]
\end{tabular}
}

\begin{document}

\title{Subaru optical observations of    
the old pulsar PSR~B0950+08\thanks{Based on data collected at Subaru Telescope, 
which is operated by the National Astronomical Observatory of Japan.}}
\author{S.V.~Zharikov\inst{1},
Yu.A.~Shibanov\inst{2},
A.B.~Koptsevich\inst{2},
N.~Kawai\inst{3,4},
Y.~Urata\inst{3,4},
V.N.~Komarova\inst{5,6},
V.V.~Sokolov\inst{5,6}, 
S.~Shibata\inst{7},
N.~Shibazaki\inst{8}}
\authorrunning{S.V.~Zharikov et al.}
\titlerunning{Subaru optical observations of the  PSR~B0950+08}
\institute{
Observatorio Astronomico Nacional SPM, Instituto de Astronomia, UNAM, Ensenada, BC, Mexico \\
zhar@astrosen.unam.mx
\and
Ioffe Physical Technical Institute, Politekhnicheskaya 26,
St. Petersburg, 194021, Russia \\
shib@stella.ioffe.rssi.ru, kopts@astro.ioffe.rssi.ru
\and
Department of Physics, Tokyo Institute of Technology, 2-12-1 Oookayama, Meguro-ku, Tokyo 152-8551, Japan 
\and 
RIKEN (Institute of Physical and Chemical Research), 2-1 Hirosawa, 
Wako, Saitama 351-0198, Japan  
\\
nkawai@hp.phys.titech.ac.jp, urata@crab.riken.go.jp
\and
Special Astrophysical Observatory of RAS,
Karachai-Cherkessia, Nizhnij Arkhyz, 357147, Russia 
\and 
Isaac Newton Institute of Chile, SAO Branch, Russia, vkom@sao.ru, sokolov@sao.ru
\and
Department of Physics, Yamagata University, Yamagata 990-8560, Japan,
shibata@sci.kj.yamagata-u.ac.jp 
\and
Department of Physics, Rikkyo University, Nishi-Ikebukuro, Tokyo 171-8501, Japan,
shibazak@rikkyo.ac.jp 
}
\offprints{S.~Zharikov}

\date{Received, accepted} 

\abstract{
We report the B band  optical observations of an old (${\rm\tau
\sim 17.5}$ Myr)  radiopulsar \psr \ obtained with  the Suprime-Cam at
the Subaru telescope.  We detected a faint object, B=27$\fm07\pm$0.16.
Within our  astrometrical accuracy
it coincides with the radio position of  the pulsar and
with the object detected earlier by Pavlov et al. (\cite{Pavlov}) in UV with
the   \hst/FOC/F130LP.   The   positional  coincidence   and  spectral
properties of the  object suggest that it is 
the optical
counterpart of PSR B0950+08. 
Its flux 
in  the  B band 
is two times higher 
than one  would expect from  the suggested earlier 
Rayleigh-Jeans interpretation  of the only available   
\hst\ observations in the adjacent F130LP band.  Based on the B and 
F130LP photometry 
of the suggested counterpart 
and  on the available X-ray data  we argue in favour 
of nonthermal origin  of the broad-band optical  spectrum of  \psr, 
as it is observed for the optical emission of the younger, middle-aged 
pulsars PSR  B0656+14 and  Geminga.  At the  same time,   
the optical efficiency 
of  \psr, 
estimated
from  
its
spin-down power 
and the detected optical flux,
is  by several orders  of magnitude higher  than for
these 
pulsars, and  comparable with that for the  much younger and more
energetic Crab  pulsar. We cannot  exclude the presence of  a compact,
$\sim$  1\arcsec,  faint pulsar  nebula  around  \psr,\  elongated
perpendicular to the  vector of its proper motion, unless  it is not a
projection  of  a  faint  extended  object  
on
the  pulsar  position.
 }
\maketitle
\section{Introduction}

PSR  B0950+08  is the fourth  
in  the  list of the first radio  pulsars
discovered in 1968  (e.g., \cite{Bell}).  Since that time  it has been
extensively studied  in a  wide radio frequency  range, from  0.102 to
10.55 GHz.  The pulsar parameters (period $P$ and its derivative $\dot
P$,  age  $\tau$,  magnetic  field  $B$,  dispersion  measure  $D\!M$,
spin-down  luminosity  $\dot  E$,  distance  $d$,  position  $\alpha$,
$\delta$,  proper  motion  $\mu_\alpha$,  $\mu_\delta$,  and  parallax
$\pi$) are shown  in Table \ref{t:par}.  
So far this is the oldest
pulsar  among ordinary  pulsars 
detected outside  the radio
range.

\begin{table*}[t]
\caption{
Parameters of   \psr\ (from \cite{Taylor}, unless specified otherwise). }
\begin{tabular}{ccccccccccc}
\hline\hline
\multicolumn{5}{c}{Observed}&&\multicolumn{5}{c}{Derived} \widerul\\
\cline{1-6}\cline{8-11}
$P$	& $\dot P$   & $D\!M$       & $\alpha_{2000}$, $\delta_{2000}^{a,b}$                             & $\mu_\alpha$, $\mu_\delta^b$     & $\pi^b$  && $\tau$ & $B$          & $\dot E$    
        & $d^{b,c}$     \widerul \\
ms      & $10^{-15}$ & cm$^{-3}$ pc &                                                                    & mas yr$^{-1}$                    & mas    && Myr    & G  & erg $s^{-1}$
        & pc       \\
\hline
253     & 0.229      & 2.97         & \twolines{$9^h53^m09\fs313(3)^d$}{7$^\circ$55\arcmin36\farcs08(4)} & \twolines{$-$2.09(8)}{29.46(7)}  & 3.82(7) && 17.5   & $0.24\times10^{12}$         & $5.6 \times 
10^{32}$   & $262(5)$ \widerul\\
\hline
\end{tabular}
\begin{tabular}{ll}
$^a$\ The  position at the epoch of the Subaru observations, MJD 51930. 
  & $^d$\ The numbers in parentheses are uncertainties \widerul\\
$^b$\ See Brisken et al.  (\cite{Brisken2}) &\phantom{$^d$}\ referring to the last significant digit quoted, \\
$^c$\ Parallax based distance &\phantom{$^d$}\ e.g., $1.234(56)=1.234\pm 0.056$ \\
\end{tabular}
\label{t:par}
\end{table*}

The X-ray counterpart of PSR B0950+08 was first found by \cite{Seward}
and  \cite{Cordova}  with  the  Einstein observatory.  Later,  it  was
observed with  the \rosat\ by Manning and Willmore  (\cite{Manning}), who analyzed  the soft X-ray
spectrum  in  (0.08--2.4)  keV  range. The  blackbody  (hereafter  BB)
spectral fit  gave a  temperature $T_{\rm BB}  = (2.1 \pm  0.6) \times
10^6  $~K (90\%  confidence limits)  and a  very small  radius  of the
emitting  region  $R_{\rm BB}  \le  20$~m.   The  pulsar distance  was
assumed  to  be  of  about  130  pc  based  on  early  radio  parallax
measurements by \cite{Gwinn} with  the VLBI.  The power law (hereafter
PL, $F_\nu\propto\nu^{-\alpha_\nu}$) spectral fit was also acceptable,
but, owing  to 
the 
poor count statistics  for this faint  X-ray object, it
resulted  in   a  very  uncertain   spectral  index  ${\rm   -0.2  \le
\alpha_{\nu} \le 3.1}$.  Some evidence, at 99\% significance level, of
the pulsed X-ray  emission with pulsed fraction ${\rm  0.34 \pm 0.18}$
was  found  by Wang and Halpern (\cite{Wang})  using  archival  \asca\  observations  in
(0.5--5) keV range.   The \asca\ spectral data were  also equally well
fitted  either by  the BB  with a  
temperature  $T_{\rm  BB} =
(5.70\pm0.63)\times 10^6 $~K, 
higher
than that in the \rosat\ range, and with
$R_{\rm  BB}   \sim  14$~m,  or   by  the  PL  with   $\alpha_{\nu}  =
0.40\pm0.22$.  In accordance with NS cooling theories, the temperature
of the  whole surface of  this relatively old  NS should be  less than
$10^5$~K (e.g.,  \cite{Yakovlev}). This is much less  than observed by
both the \rosat\ and \asca.  The  high temperature and small emitting area
inferred from the BB fit can be explained by the presence of small hot
polar caps  at the NS surface  produced by the  impact of relativistic
particles from  the pulsar  magnetosphere, see, e.g.,  Wang and Halpern (\cite{Wang}) and
references
therein.

An optical counterpart of  PSR B0950+08 was suggested by Pavlov et al. (\cite{Pavlov})
based  on observations of the  pulsar field with  the \hst/FOC with
the  long-pass F130LP  filter ($\lambda\lambda  = 2310-4530$  \AA).  A
faint ($m_{\rm F130LP}=27$\fm$1$) point-like object was found with the
1\farcs85 offset from the pulsar radio position.  The offset was later
revised and decreased to $\simeq 1$\arcsec\   by Pavlov (\cite{Pavlov1}).  
If this is 
a
pulsar,
it  is the  faintest 
pulsar
ever detected in the optical. For comparison, the visual
magnitude  of the young  Crab pulsar,  which is  about ten  times more
distant but  $1.75\times 10^4$ times younger, is  16$\fm65$ (e.g., see
the review  by \cite{Mignani2}).   

Assuming that the  detected optical
object is the  pulsar, Pavlov et al. (\cite{Pavlov}) showed that the  
extension of the
\rosat\ and  \asca\ BB  fits into the  UV-optical range 
gives  a flux lower than observed by  3--4 orders of magnitude.   
This excludes
thermal radiation from  the pulsar polar caps as  a possible source of
the optical  radiation. The assumption that the  detected optical flux
is due to thermal  emission from the entire surface of a  NS with a BB
radius $R_{\rm BB}=13$~km yielded  the surface temperature $T_{\rm BB}
\sim 7 \times 10^4$~K at ${\rm  d=130}$~pc. This is still too high for
the $\sim 1.75  \times10^7$~yr cooling NS and can  be only explained by
some   reheating   mechanism   operating   inside  the   star   (e.g.,
\cite{Miralles}).  On  the other  hand, the optical  flux would  be in
agreement with  the extension of the  PL X-ray fit if  the index ${\rm
\alpha_{\nu}}$ lies  within the 0.26--0.35 range.  It  is in agreement
with  the \rosat\  and \asca\  data but  needs 
a confirmation  by deeper
observations in X-rays and by  the detection of the counterpart in    
other optical bands.

Possible  detection of  PSR  B0950+08 in  the R  band with  the 6m
telescope BTA  has been  reported by \cite{Sokolov}  and \cite{Kurt1}.
An  object  with  $R=25\fm4\pm0.3$  was marginally  ($S/N  \simeq 3$)
detected  with  a  short  exposure  in  poor  seeing  conditions.   If
confirmed,  this  detection  suggests  that  the pulsar  may  be  much
brighter in  the optical  and may  have a very  steep increase  of the
spectrum towards  longer wavelengths  than one would expect  from the
detection in the near UV.

In this paper, we report  the observations of the PSR~B0950+08 field in
the B  band with the Subaru  telescope.  We analyze  our data together
with the available  optical-UV data from the \hst,  and with the X-ray
data  from the \asca\  and \rosat,\  making use  of the  recent  much more
accurate radio measurements of the pulsar proper motion, parallax, and
distance  with  the   VLBA  by  
Brisken et al. (\cite{Brisken2}).
Observations and data reduction are  described in Sec.~2. In Sec.~3 we
present the astrometrical referencing and photometry, and in Sec.~4 we
discuss the results and their implications.

\begin{table}[b]
\caption{Log of observations of the pulsar in the B band.}
\begin{tabular}{cccccc}
\hline\hline
Exposure	&       UT  	& Duration	   & Airmass & Seeing    \\ 
number		& 21 Jan 2001	&   sec            &         &  arcsec   \\ \hline 
1		& 12:41     	&   600            & 1.037   &  0.65     \\
3		& 13:21     	&   600            & 1.085   &  0.66     \\ 
4		& 13:54     	&   600            & 1.129   &  0.69     \\ 
5		& 14:08     	&   600            & 1.164   &  0.72     \\ 
6		& 14:22     	&   600            & 1.205   &  0.73     \\ 
8		& 14:50     	&   600            & 1.313   &  0.75     \\
 \hline
\end{tabular}
\label{t:log}
\end{table}

\section{Observations and data reduction}

The field  of  \psr \  was observed on  Jan 21, 2001 with  the wide
field camera Suprime-Cam at the primary focus of the Subaru telescope.
The Suprime-Cam  (\cite{Miyazaki}) is  equipped with ten  MIT/LL $2048
\times  4096$  CCDs arranged  in  a  $5\times  2$ pattern  to  provide
$34^{\prime} \times 27^{\prime}$  FOV with a pixel size  on the sky of
$0.201^{\prime\prime}  \times 0.201^{\prime\prime}$.   The  pulsar was
exposed in one  of the CCD chips, {\tt  si006s}, with the {\tt gain}$=$1.17.
Nine 600 s  exposures were taken using the B  filter with the bandpass
close    to   the    Johnson   system.  The   mean   seeing    was   about
$0.7^{\prime\prime}$.  The  2nd, 7th,  and 9th exposures  were removed
from  further consideration  because  of 
the
problems  with the  telescope
guiding system.  The observational  conditions for the rest six 
exposures are listed in Table \ref{t:log}.

\begin{figure*}
\setlength{\unitlength}{1mm}
\begin{picture}(179,177)(2,0)
\put 	(-30,92)					{\includegraphics[width=120mm,bb=15 90 450 390,clip]{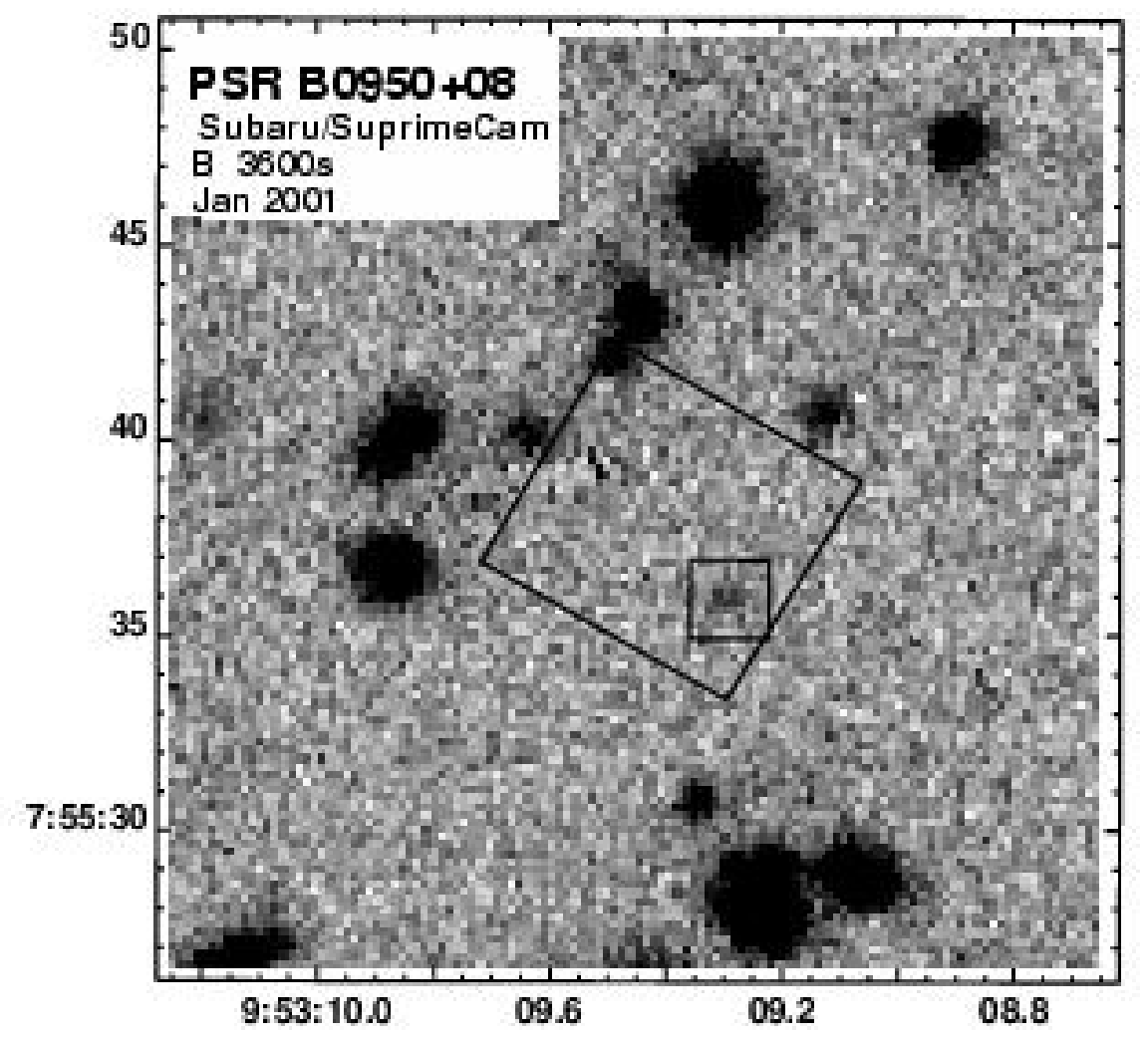}}
\put 	(-20,0)					{\includegraphics[width=110mm]{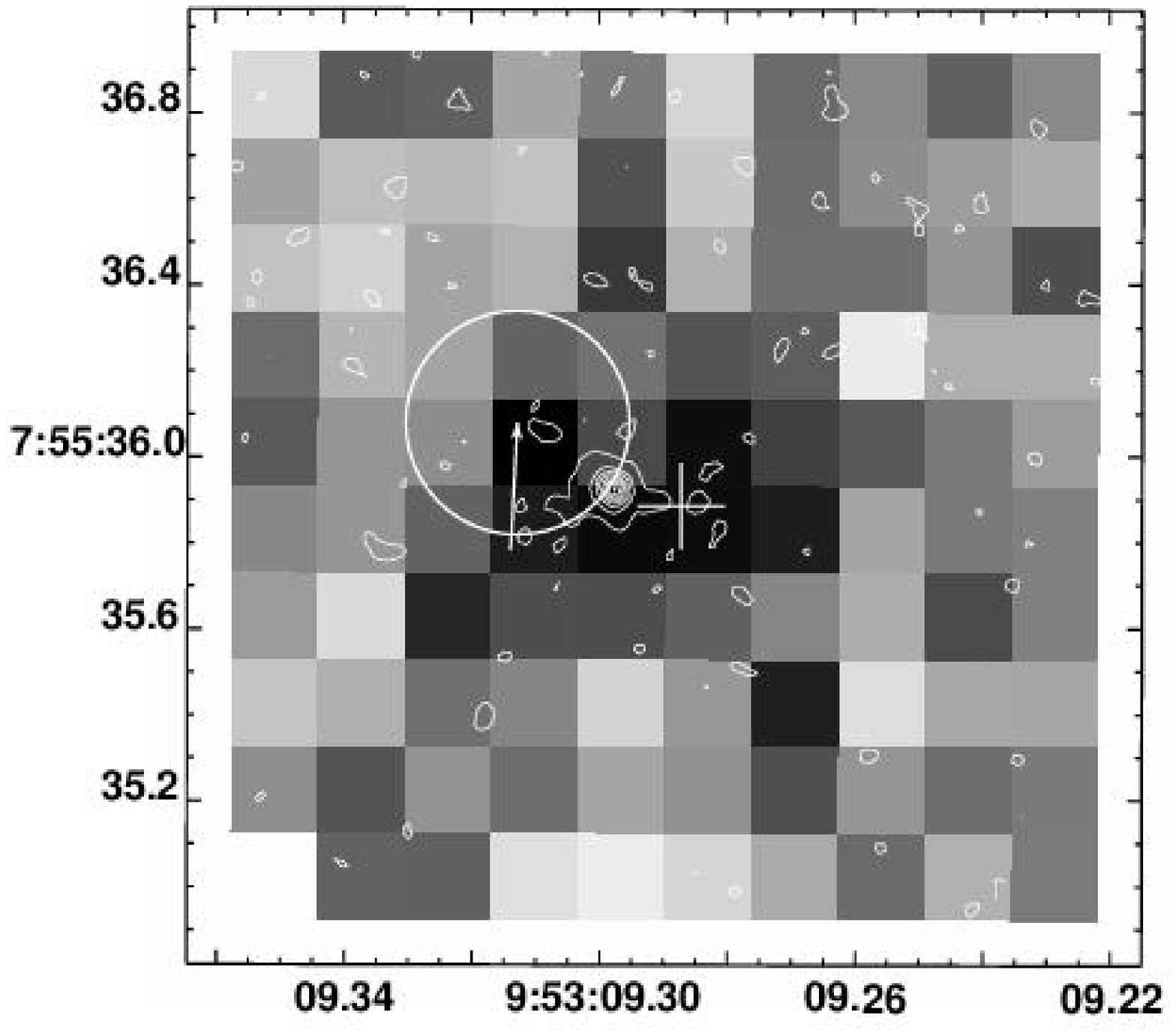}}
\put	(90,90)				{\includegraphics{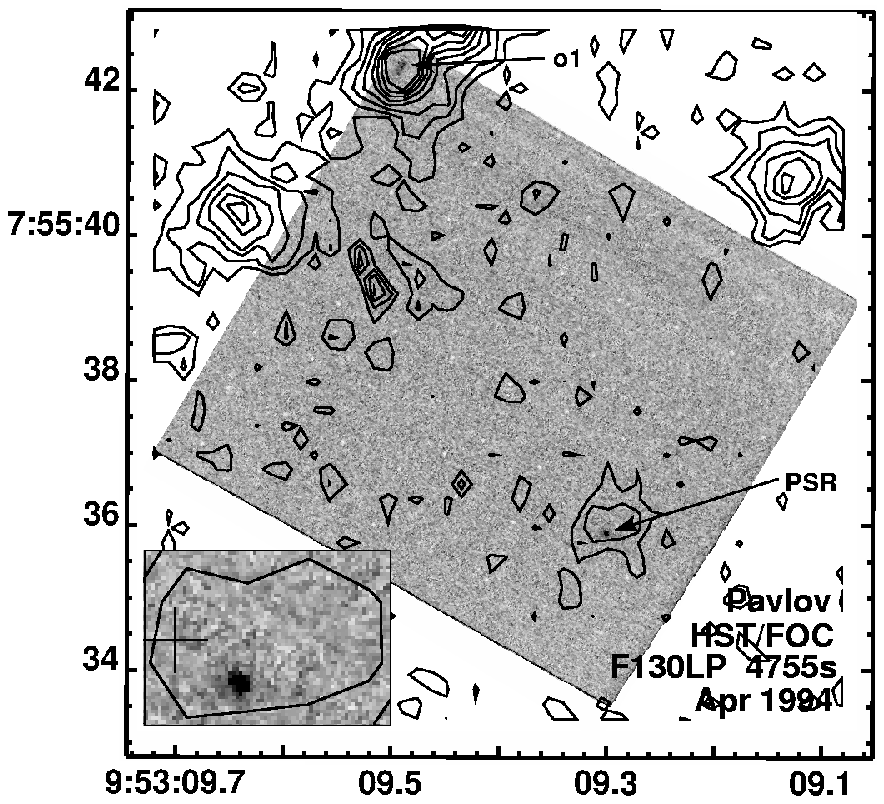}}
\put	(93,0)					{\parbox[b]{88mm}{\caption{%
{\bfseries\itshape Top Left (\ref{f:ima}a):}
  24\asec$\times$24\asec\ fragment of the Subaru 2001 image of
  the PSR B0950+08 field in the B band. 
  Larger and smaller boxes show the FOV of the \hst\  
1994 observations (see the {\it right\/} panel)  
  and the borders of the {\it bottom\/} panel containing the pulsar, respectively.
{\bfseries\itshape Right (\ref{f:ima}b):}
  7\farcs4$\times$7\farcs4 \hst\ 1994 
  image of the pulsar field. 
  The pulsar counterpart and an extended object o1, both visible in the \hst\ and Subaru images,  are marked.
  The \hst\ image is aligned with the Subaru image using o1 as a reference.
  The Subaru contour map is overlaid. The pulsar close vicinity, $\sim0.6$\arcsec,  
  with the closest Subaru contour and the radioposition 
at the epoch of the FOC observations marked by "+",   
is zoomed in the inset. 
{\bfseries\itshape Bottom (\ref{f:ima}c):}
  2\asec$\times$2\asec\ neighbourhood of the pulsar in the B band.
  The \hst\ contour map is overlaid. It is shifted for the pulsar proper motion 
  for 6.75 yr passed between 
  the  observations.  The arrow points out the radio position of the pulsar at the epoch of the Subaru observations, 
  its length corresponds to the pulsar shift due to the proper motion for 10 yr.
  The circle shows 1$\sigma$ 
uncertainty
  of the pulsar position in the Subaru image;  "+" marks the center of the Gaussian 
  fitted to the pulsar counterpart profile in this image. 
\label{f:ima} 
}}}
\end{picture}
\end{figure*}

The   standard  data   reduction,  including   bias   subtraction  and
flatfielding, was  performed making use  of the {\tt ccdproc}  task of
the {\tt  IRAF} software.  To  combine the individual  dithered images
and to  get rid of cosmic  rays, as well as  of a trace  of an unknown
minor  object or  a  dust clump  with  an orbit  close  to the  Earth,
crossing  our image  E--W  approximately 8\asec\  away  of the  pulsar
position,   we   made  use   of   the   {\tt 
IRAF
dither}  package.
A fragment of the combined image of the pulsar field
with  the  total  integrated exposure  time  of  3600  s is  shown  in
Fig.~\ref{f:ima}a\footnote{Image is available in FITS format at 
\href{http://www.ioffe.rssi.ru/astro/NSG/obs/0950-subaru}
{http://www.ioffe.rssi.ru/astro/NSG/obs/0950-subaru}.}.

\section{Astrometrical referencing and photometry}
 
\subsection{Astrometry} 

The  radio  position of  \psr\ at the epoch of the observations   
(see Table~\ref{t:par}) was determined using the VLA observations 
by \cite{Fomalont} and the most recent measurements of the pulsar proper 
motion by Brisken et al.  (\cite{Brisken2}).

Astrometric  referencing of  our  image was  made  using positions  of
five\footnote{U0975\_06358239,  U0975\_06357626,  U0975\_06357575,
U0975\_06358752, U0975\_06358583.}   of 
nine  reference stars from
the      USNO     A-1.0      catalogue     visible      within     the
$\sim6\arcmin\times13\arcmin$  frame of  the CCD  chip  containing the
pulsar.   The  PSFs  of these  stars  are  not  corrupted by  the  CCD
oversaturation effects.   We used {\tt IRAF}  tasks {\tt ccmap/cctran}
for  the astrometric transformation  of the  image.  
RMS
errors of the astrometric fit are $0\farcs086$ and $0\farcs090$ for RA
and Dec, respectively, 
whereas the
residuals  for all stars are $< 0\farcs13$.
All  of  them  are  less  than the  nominal  USNO  catalogue  accuracy
0\farcs24.   Combining the  
RMS
errors of  the  fit, the  USNO
catalogue    accuracy   and   the    radio   position    errors   (see
Table~\ref{t:par})  we  estimate  total  uncertainties of  the  pulsar
position,  as well  as the  astrometric referencing  accuracy  for the
whole Subaru image, as 0\farcs26 in both RA and Dec\footnote{There are
no reference stars south of the pulsar within our frame and 
thus
we cannot
exclude  a systematic  shift in the N--S direction.}.

\subsection{Identification and morphology of the candidate to the pulsar counterpart}

As seen  from Fig.~\ref{f:ima}, the  optical counterpart of  \psr\
can only  be identified with a  faint, ${\rm S/N  \simeq 7}$, isolated
object clearly visible near  the center of the 24\asec$\times$24\asec\
fragment of the Subaru image (Fig.~\ref{f:ima}a).  The object overlaps
with
the error circle  of the pulsar radio position  with radius 0\farcs26,
corresponding   to  $1\sigma$   
uncertainty
of   our  astrometrical
referencing (Fig.~\ref{f:ima}c).  The source  profile of the object is
elongated E--W,  with FWHMs  $\sim 1\farcs0$ and  $\sim0\farcs7$ along
E--W and S--N  directions, respectively.  Owing to the  low S/N ratio,
we  do  not  resolve   any  point-like  structure  within  the  source
profile. 
Formal gaussian  fitting of the whole object profile
yields   the   coordinates   $\alpha_{2000}  =   9^h53^m09\fs288(17)$,
$\delta_{2000} = 7^\circ$55\arcmin35\farcs89(26).  The offset from the
radio position  is 0\farcs43 W--S.   It is within $1.75\sigma$  of the
astrometrical accuracy  and can be considered as  negligible given our
seeing and S/N values.

To compare our image with  the previous optical-UV observations of the
pulsar field by  Pavlov et al.  (\cite{Pavlov}) we reanalyzed astrometrical referencing
of the \hst/FOC  image retrieved from the \hst\  archive.  The rotated
box   in   Fig.~\ref{f:ima}a  borders   the   FOV   of  the   \hst/FOC
observations.  Pavlov et al. (\cite{Pavlov}) found the  only point-like object with the
offset  $\simeq$1\arcsec  ~ (Pavlov \cite{Pavlov1})   from  the  pulsar  radio
position.  Such a large offset  makes identification of the FOC object
with the  pulsar and with  the Subaru pulsar counterpart  doubtful. We
revised the FOC  astrometry making use of  the FOC position  angle and the
only reference point visible at the  north corner of the FOC image, an
extended object  o1 (see Fig.~\ref{f:ima}b).  We assumed that o1  is a
distant  background  object  and  its  proper  motion  is  negligible.
Gaussian center coordinates of o1  were determined in the Subaru image
($\alpha_{2000}=9^h53^m09\fs489(2)$,
$\delta_{2000}=7^\circ$55\arcmin42\farcs2(1))  and  in  the FOC  image
(with  the  accuracy  $\la  0\farcs02$),  and  were  used  to  correct
reference frame of  the FOC image.  The coordinates  of the FOC pulsar
candidate in the corrected system at the epoch of the FOC observations
are                                    $\alpha_{2000}=9^h53^m09\fs298$,
$\delta_{2000}=7^\circ55\arcmin35\farcs76$. 
Given that,
the   discrepancy between the  FOC counterpart  position and the  radio position  at the
epoch  of  the  HST observations,  $\alpha_{2000}=9^h53^m09\fs314(3)$,
$\delta_{2000}=7^\circ55\arcmin35\farcs88(4)$,  marked by  "+"  in the
inset  on  Fig.~\ref{f:ima}b, is  decreased  to  0\farcs27.  This  is
comparable  to the  astrometrical referencing  accuracy of  the Subaru
image.

The  FOC image  is  presented in  Fig.~\ref{f:ima}b  in the  corrected
coordinates.  The overlaid contour map  of the Subaru image shows that
the  Subaru  and  \hst\  detected  the same  object  near  the  pulsar
position.  The  FOC contour  map in Fig.~\ref{f:ima}c  is additionally
shifted with  respect to the Subaru  image by $-$0\farcs014  in RA and
0\farcs199 in  Dec to compensate  for the pulsar proper  motion during
the    6.75   yr    interval    between   the    \hst\   and    Subaru
observations. Isophotes  on the contour maps correspond  to the levels
(in  counts) above the  background $l_n=S+n\sigma$,  where $S$  is the
mean sky value near the pulsar, $\sigma$ is the sky standard deviation
related to one pixel, $n=1,2,3,\ldots,8$ and $1,3,5,\ldots,15$ for the
Subaru and \hst\ maps, respectively. From  Fig.~\ref{f:ima}c we see
that the  Subaru and  \hst\ detected the  same object near  the pulsar
position. The  better agreement  of the Subaru  and HST  source positions 
 after the correction for  the pulsar  proper motion favours 
 this object  as the optical counterpart of \psr.

We also note that, although the \hst/FOC pulsar counterpart profile is
point-like,  the  edges  of  its  wings are  slightly  elongated  E--W
(Fig.~\ref{f:ima}c).  The  elongation directions  in both the  FOC and
Subaru images coincide and  are approximately orthogonal to the vector
of  the  proper   motion  of  the  pulsar  marked   by  the  arrow  in
Fig.~\ref{f:ima}c.  Such an orientation  may suggest the association of
the elongation with a faint torus-like structure of a possible compact
pulsar  nebula  as  detected  around young  Crab-like  pulsars  (e.g., 
Weisskopf et al.~\cite{We}). 
However, it may be  also a projection 
of a faint extended background object at the pulsar position.

To summarize,  our analysis shows  that the \hst\ and  Subaru detected
the same object. With the  allowance for the pulsar proper motion, the
offsets  from  the \psr\  radio  positions  at  the \hst\  and  Subaru
observational epochs are in the range (0\farcs3 -- 0\farcs4), which 
are
within $1.7\sigma$  error of the Subaru  astrometrical referencing and
negligible compared to our seeing of 0\farcs7.

\subsection{Photometry}

Weather conditions were stable during our observations. We derived the
atmospheric  extinction coefficient in  the B  band $k_B=  0$\fm$18 \pm
0.02$  from the  variation of  the count  rates of  four stars  in the
pulsar   field    with   airmass   during    our   observations   (see
Table~\ref{t:log}).  
Insignificant  decrease of the  extinction, within
1-$\sigma$  level,  was noticed  from  the  beginning  to the  end  of
observations.

Photometric referencing was carried out using three defocused standard
stars  from the  field PG1047  (\cite{Landolt}) with  $V\sim13$, along
with five  fainter, $V\sim  21$, unsaturated secondary  standard stars
from the  PSR B0656+14 field  (\cite{Kurt}) observed in the BRI  bands the
same night.  The derived photometric  zeropoint 
in the B band 
was
$28.28\pm0.02$.  The  instrumental magnitudes of  the detected optical
source 
were
measured for a range  of aperture radii of ($1-3$) CCD pixels
centered at the "+" in  Fig.~\ref{f:ima}c. They were corrected for the
PSF of bright  stars (some details on the method we  used can be found
in  Koptsevich et al. \cite{Koptsevich}).  
Within the measurement errors, the results for
different  apertures coincided  and a  2 pixel  radius was  adopted as
optimal.    The   magnitude    of   the   detected   optical   source,
$B=27.07\pm0.16$,   corresponds   to   the   absolute   flux   
$F_{\rm
B}=(5.97\pm0.88)\times 10^{-31}$  erg cm$^{-2}$ s$^{-1}$  Hz$^{-1}$ or
$0.0597 \pm 0.0088$~$\mu$Jy.  The  error includes statistical error of
the  instrumental magnitude measurements,  the  error of  
the zeropoint, and
an allowance for possible atmospheric extinction variations.

\begin{figure}[t]
\begin{center}
\includegraphics[width=88mm,clip]{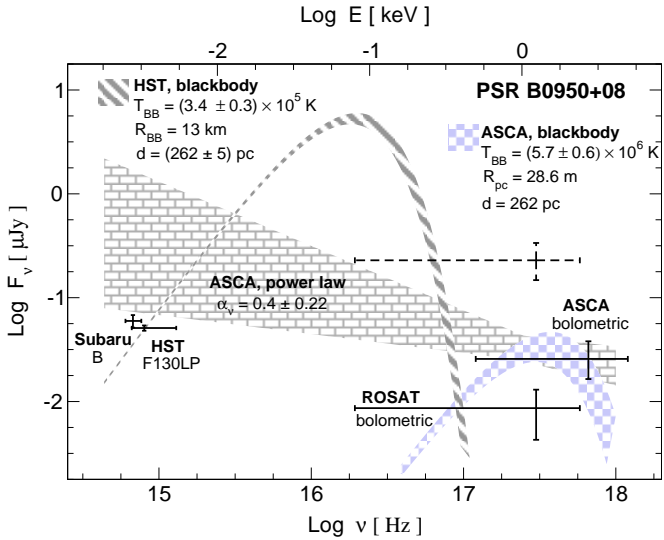}
\end{center}  
\caption{Optical and X-ray observations of \psr. Solid
crosses in the X-ray range correspond to the bolometric fluxes  derived from the BB
and PL  spectral fits of the \rosat\ (Manning and Willmore 
\cite{Manning}) 
and
\asca\  (Wang and Halpern  
\cite{Wang}) 
data in the $E=(0.08-2.4)$ and $(0.5-5)$ keV energy bands, respectively.
Square-filled belt and brick-filled region show unabsorbed spectra resulted from
the BB and PL fits of the \asca\  data. Stripe-filled belt shows unabsorbed BB fit of the \hst/FOC data,
and dashed cross shows corresponding bolometric flux in the ROSAT band. 
The fits are  extrapolated toward the X-ray and optical ranges, 
respectively, and their parameters are shown in the plot. 
The  widths of belts and region correspond to $1\sigma$ uncertainties of the fits. 
 As seen,
the optical emission is  
of nonthermal origin and apparently follows the extrapolation 
of the X-ray PL fit of the ASCA data with the spectral index 
$\alpha_{\nu} \simeq 0.18$ 
(see Sec.~\ref{s:disc} for details).}
\label{f:flux}
\end{figure}
 
\section{Discussion}

\label{s:disc}

In Fig.~\ref{f:flux} 
we combine our results with the available data from the optical and 
X-ray observations of \psr. 
In the optical range we neglect the interstellar extinction,
which is expected  to be very low for this pulsar,
$E_{B-V}\la 0.02$ (e.g., Pavlov et al. \cite{Pavlov}).
We excluded from our consideration the observations in 
the R band by \cite{Kurt1} because of very low signal-to-noise ratio.  
In the X-ray range 
we show 
unabsorbed fluxes resulted  from  the BB and PL 
spectral fits of the \rosat\  (Manning and Willmore  
\cite{Manning}) 
and \asca\  
(Wang and Halpern  
 \cite{Wang}) 
data. The fits are rescaled to the new distance 
value $d=262\pm5$ pc  measured by Brisken et al.  (\cite{Brisken2}).       
  
Within the errors the flux 
$F_{\rm B}$ coincides with the value 
$F_{\rm F130LP}=0.051\pm0.003$~$\mu$Jy
measured by  Pavlov et al. (\cite{Pavlov}) with the \hst/FOC
in the F130LP band. The B band 
($\lambda\lambda \sim 3944-4952$~\AA)
considerably 
overlaps with the F130LP band   
($\lambda\lambda \sim 2310-4530$~\AA). 
However, their  pivot  wavelengths $\lambda^0_{\rm B}=4448$~\AA\  and 
$\lambda^0_{\rm F130LP}=3750$~\AA\ are different.  
This difference and  
close flux values in two bands  
suggest a flat spectrum of the object in  
a wide  $\lambda\lambda \sim 2300-5000$~\AA\ spectral range.  
Along with the positional coincidence,
we consider this  flat spectrum, which is typical for 
 pulsar optical spectra 
(e.g., Koptsevich et al.  
\cite{Koptsevich};     
Mignani \& Caraveo \cite{Mignani3} ),  
as an 
additional
argument 
in favour of the detection of the \psr\ optical counterpart.

If the detected object is the pulsar,  its flux in the B band does not
follow the  Rayleigh-Jeans law  suggested by  Pavlov et al. (\cite{Pavlov})  to explain
the optical-UV radiation in the  only F130LP band as a low temperature
thermal  emission from the  entire surface  of a  reheating/cooling NS
with $R_{\infty}=13$ km.   In that case in the B  band we would  detect 
twice smaller flux  than  measured.  This is well outside
the  uncertainties and  suggests  a nonthermal  origin  of the  pulsar
emission at least in the B band.

\begin{table*}
\caption{
Parameters of the radio pulsars and the efficiencies of their optical emission $\eta_B$ in the B band.
}
\begin{tabular}{lccccccc}
\hline\hline
Source  	& $\log\tau$	& $d$    					& $\log\dot{E}$	& $B$ 		& $M_{\rm B}$		& $\log L_{\rm B}$		& $\log\eta_{\rm B}$ \\ 
        	&  yr		& kpc      					& erg s$^{-1}$	& mag 		& mag 			& erg s$^{-1}$			&           \\ 
\hline
Crab		&  3.1       	& 2.0(1)           				& 38.65		& $15.25(7)^1$	& 3.74(13)   		& 33.23(5)			& $-$5.42(5)   \\ 
B0540$-$69	&  3.2       	& $50^{+5}_{-0.6}$          			& 38.17		& $22.0(3)^2$	& 3.48(37)		& 33.47(15)			& $-$4.7(2) \\
Vela		&  4.1       	& ${0.294^{+0.076}_{-0.050}}^a$			& 36.84		& $23.7^(3)^3$	& $16.4^{+0.5}_{-0.8}$	& 28.3(3)			& $-$8.5(3)   \\
B0656$+$14	&  5.0       	& ${0.5^{+0.26}_{-0.3}}^b$           	& 34.58		& $25.15(13)^4$	& $16.8^{+2.1}_{-1.0}$	& $28.2^{+0.4}_{-0.9}$		& $-6.4^{+0.4}_{-0.9}$\\
Geminga		&  5.5       	& ${0.153^{+0.059}_{-0.034}}^{c}$ 		& 34.51		& $25.7(3)^5$	& $19.8^{+0.8}_{-1.0}$	& $26.95^{+0.16}_{-0.10}$	& $-7.56^{+0.16}_{-0.10}$  \\
B1929$+$10	&  6.5       	& $0.331(10)^{d}$ 	& 33.59		& $ \geq26.2^6$	& ${20.0^{+0.2}_{-0.2}}^{**}$		& ${27.26^{+0.2}_{-0.3}}^{**}$		& ${-6.3^{+0.2}_{-0.3}}^{**}$ \\
B0950$+$08	&  7.2      	& 0.262(5)$^{d}$				& 32.75		& $27.07(16)^7$	& 19.98(19)		& 26.88(8)			& $-$5.87(8)  \\ \hline
 \hline \\
\end{tabular}
\label{t:eff}
\begin{tabular}{l}
The B band magnitudes are from: \\
$^1$ \cite{Percival};
$^2$ \cite{Middleditch} (spectroscopic data by \cite{Hill} give $\sim 1^{m}.5$ smaller magnitude); \\
$^3$ \cite{Nasuti};
$^4$ Koptsevich et al. \cite{Koptsevich};
$^5$ \cite{Bignami}; 
$^6$  Pavlov et al. (\cite{Pavlov}) (3$\sigma$ upper limit); \\
$^7$ This work. 
The distances $d$ are from : \\
$^a$ \cite{Caraveo};
$^b$ \cite{Taylor}, \cite{Finley}, \cite{Anderson}; 
$^c$ \cite{Caraveo2} \\
$^d$ Brisken et al.  (\cite{Brisken2});
 \\
$^{**}$ The estimates are based on the counterpart detection in the adjacent F130LP band by Pavlov et al.  (\cite{Pavlov}), see Sec.~\ref{s:disc}.\\
Uncertainties of $M_B$, $L_B $ and $\eta_B$  include uncertainties of the optical flux and distance measurements.   
\end{tabular} 
\end{table*}

If the flux in the F130LP band is still dominated 
by  the Rayleigh-Jeans tail, we would obtain  
$T_{\rm BB}=(3.4\pm0.3)\times 10^5$  K, 
which is a factor of four-five higher than found by  Pavlov et al. (\cite{Pavlov}) 
mainly owing to the change of
the  distance to  \psr \ from 130~pc to 262~pc 
(see the stripe-filled BB belt  
crossing the F130LP band and extended to X-rays in Fig.~\ref{f:flux}).
A NS with such a hot surface would  produce a   
bolometric flux 
$ \sim (0.8-1.9)\times 10^{-12}$ erg cm$^{-2}$ s$^{-1}$
in the ($0.08-2.4$) keV  band 
(see a big dashed cross in  Fig.~\ref{f:flux}).    
It is well above the value
 $(2.4 - 7.3)\times 10^{-14}$  erg cm$^{-2}$ s$^{-1}$,    
measured by Manning and Willmore (\cite{Manning})  
with the \rosat\ under the assumption of the BB spectrum 
of the detected X-ray radiation (marked by a big 
cross below the dashed one in Fig.~\ref{f:flux}). 
This means that the whole surface of \psr\
is actually much cooler and its emission 
cannot dominate  in the F130LP band.  
Thermal emission from hotter, $T_{\rm BB}=(5.7 \pm0.6) \times 10^6$~K, 
but much smaller polar caps of the pulsar with $R_{\rm pc} \sim 28.6$~m, 
which may explain the detected 
X-ray radiation (Wang and Halpern    
\cite{Wang}), 
can  hardly be  visible 
in the optical range also  
because of very small areas of the caps  inferred from 
the BB fit of the X-ray data  
(see square-filled BB belt in Fig.~\ref{f:flux}).

For the above reasons it is most likely that the optical radiation 
of  \psr \ is completely dominated by nonthermal emission produced 
in the magnetosphere of the rapidly rotating NS, as it is believed to be for
young and well studied pulsars like the Crab and PSR  B0540$-$69.
Within large uncertainties of the 
available
X-ray data for \psr \ the PL with the spectral 
index $\alpha_{\nu} \simeq 0.18$ matches both the X-ray and optical 
fluxes  including the B and F130LP bands.  This value is consistent 
with what was obtained by Wang and Halpern (\cite{Wang}) from the  analysis of 
the  \asca\ X-ray data, $F_{\rm ASCA}^{\rm PL} = 2.77~(+1.33,-1.05)\times 10^{-13}$ 
erg cm$^{-2}$ s$^{-1}$, ~$\alpha_{\nu}=0.40 \pm 0.22$   
in (0.5 -- 5) keV energy 
range (see the ASCA  
cross and the brick-filled region in Fig.~\ref{f:flux}).
The PL fit of the X-ray data is more 
preferable since the BB fit, which is statistically 
also acceptable,  implies by an order of magnitude smaller emitting area 
than it is predicted by standard models 
of hot polar caps at the pulsar surface (e.g., \cite{Arons} ). 
The inferred spectral index differs from   
$\alpha_{\nu} \sim 1$ which is typical for nonthermal soft 
X-ray radiation of  most rotation-powered NSs   
(e.g., \cite{Becker}). 
However, it can be as low as 0.4 for middle-aged pulsars 
(Koptsevich et al.  
\cite{Koptsevich}) 
and we cannot exclude  a decrease  
of the slope of the \psr \  spectrum  towards the optical range 
as it is seen in the case  of the Crab pulsar (e.g., 
Crusius-W\"{a}tzel et al.   
\cite{Crusius}).
More X-ray and optical data are needed to check the spectral shape  for \psr. 
With  new data it would be also useful to perform  the BB $+$ PL, 
and/or NS atmosphere $+$ PL  fits (e.g., Zavlin et al.  
\cite{Zavlin}) 
 to better  constrain  
the parameters of the nonthermal and thermal spectral components 
from the pulsar magnetosphere and polar caps  
and to estimate their contribution to the pulsar emission 
in different spectral bands.

\begin{figure}[t]
\begin{center}
\includegraphics[width=88mm,clip]{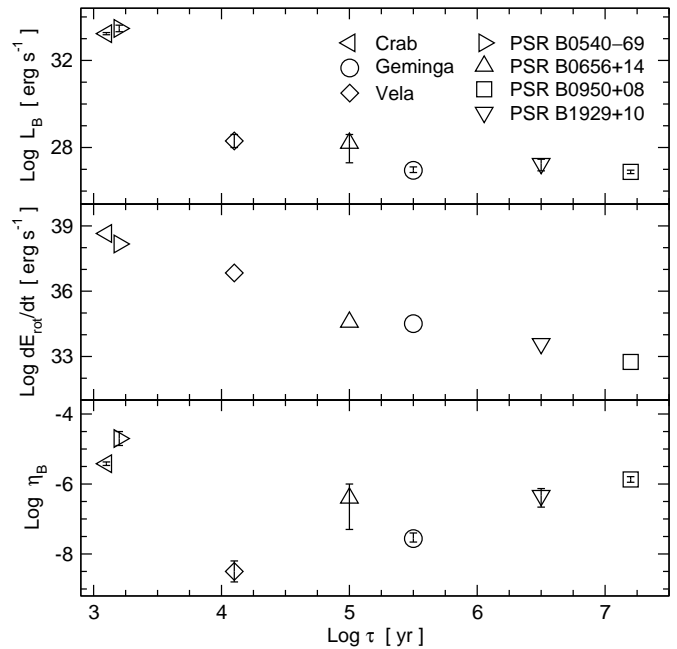}
\end{center}  
\caption{The  optical luminosity  $L_{\rm B}$  ({\it  top}), spin-down
power $\dot{E}$  ({\it middle}) and  optical efficiency in the  B band
$\eta_B$ ({\it bottom}) for all  radio pulsars detected in the optical
range as functions of pulsar age $\tau$ (see also Table~\ref{t:eff}).}
\label{f:eff}
\end{figure}
     
If the detected object is \psr, the pulsar optical luminosity in the B band,
assuming  isotropy   of  the   emission,  is  $L_{\rm   B}  =   4  \pi
d^2F_\nu\Delta\nu = (7.6\pm1.2)\times10^{26} d^2_{262}$~ erg s$^{-1}$,
where $  d_{262}=d/(262$~pc) is the normalized distance.  As seen from
Table~\ref{t:eff} and Fig.~\ref{f:eff} (top panel), where we collected
the data in the B band  for all pulsars detected in the optical range,
$L_{\rm  B}$ of  \psr\ is  comparable with  the luminosity  of the
middle-aged  Geminga pulsar  but  by almost  six  orders of  magnitude
smaller than 
$L_{\rm  B}$ of
much younger and energetic Crab  pulsar.  These data
indicate  that, starting approximately  from the  age of  Geminga, the
decrease of the optical luminosity  with the pulsar age becomes slower
than  at early  stages  of  the pulsar  evolution  or even  disappears
completely, while the spin-down  luminosity continues to decrease with
approximately the same rate (Fig.~\ref{f:eff}, middle panel). This means
that for unknown reasons mechanisms  of the optical emission may start
to be more efficient again after $\sim 10^6$~yr age.  The ratio of the
optical  luminosity  to the  pulsar  spin-down luminosity,  $\eta_{\rm
B}=L_{\rm   B}/\dot{E}$,  which   can   be  considered   as  a   rough
characteristic  of  this efficiency,  is  plotted in  Fig.~\ref{f:eff}
(bottom panel).  It  is seen that optical photons  are produced by the
old  PSR B0950+08 with  almost the  same efficiency  as by the young Crab
pulsar.   In  this respect  it  is not  surprising  that  we see  some
indications  of the  presence of  a compact  pulsar nebula  around 
\psr. Manning and Willmore \cite{Manning} discuss 
a compact synchrotron nebula,  which is unresolved with the \rosat, 
as the most likely explanation of the observed  
X-ray spectrum of the pulsar. The estimated sizes of the possible 
X-ray nebula $\sim (6\arcsec - 1\arcmin)$ are much larger than 
$\sim 1\arcsec$  we  see in our optical 
images. But the difference in the X-ray and optical sizes is typical for known compact pulsar  nebulae
where only internal brightest structures of a nebula are visible in the optical 
range (e.g., Weisskopf et al.~\cite{We}). 

The efficiency derived by  Pavlov et al. (\cite{Pavlov}) 
in the F130LP band for
another old, $\sim 3.1\times 10^{6}$ yr, PSR B1929+10 is about $ (3-8)
\times 10^{-7}d^2_{170}$. This pulsar 
has not yet been detected in the
adjacent B band, but we can assume that its flux in B is close to that
in F130LP,  as it is  for   PSR  B0950+08.  Scaling by the  factor $
\Delta  \nu_{\rm B}/\Delta  \nu_{\rm F130LP}  = 0.24$  and by  the new
radio parallax based distance  $d=331$~pc (Brisken et al. \cite{Brisken2}) yields
$\eta_{\rm  B} \sim  (2.7-7.3) \times  10^{-7}d^2_{331}$.
With this value PSR B1929+10 occupies an intermediate  position  
at the rising part of the dependence of $\eta_{\rm B} 
(\tau)$, between  
Geminga and \psr,\ as it 
is
expected from its age.  
Although our consideration of the efficiency evolution  
is based  on the data obtained  only in the B
band, it appears to be qualitatively valid for the whole optical range
since the  broad-band spectra of  all pulsars are  almost flat. Hence,
the differences  of the bolometric  optical fluxes of  various pulsars
should not be affected  strongly by insignificant differences of their
spectral slopes in this range.

Physical reasons for such a high increase of the optical efficiency at
late stages of the pulsar evolution  are not quite clear.  We can only
note   that   the   efficiency   for  
gamma-ray   
radiation,
$\eta_{\gamma}=L_{\gamma}/\dot{E}$,  also   appears  to  be  generally
higher for older pulsars  (e.g., \cite{Thompson}). It is difficult to
estimate statistical significances of these facts and their possible
correlation  since the numbers of pulsars currently detected in  the optical   
and 
gamma-rays
are too small (about ten only). 
Morever, not all known 
gamma-ray pulsars
are detected in the optical range 
(and {\it vice versa}). However, the observed tendency in both  ranges 
seems to be interesting and can  hardly be ignored. 
It is obvious that in the both wavelength  ranges  the radiation 
is  nonthermal  and   originates  in magnetospheres  of rapidly rotating  NSs. 
Thus,  the increase  of the
efficiencies  in two  very  different ranges  may  reflect an  overall
increase  of the  magnetospheric  activity with  the  NS spin-down.  A
global  electrodynamic  model of  the  pulsar  magnetosphere with  the
activity caused by the magnetic field-aligned potential drop producing
electron-positron  pairs   in  the  magnetic  polar   regions  of  the
magnetosphere predicts  that the  efficiency should increase  for high
energy photons $ \propto  P^2$ (\cite{Shibata}).  However, the optical
data, particularly  for young pulsars,  do not follow  this dependence
and detailed  studies of  the electrodynamics and  radiation processes
still have to be done to explain the efficiency evolution in different
spectral  ranges.  

Further  observations of  the
candidates to the optical counterparts of the old pulsars PSR B0950+08
and  PSR B1929+10  in different  spectral bands  would be  very  useful to
resolve the efficiency problem and to better understanding  
the nature of the optical emission of  pulsars and  the evolution of  this emission  with pulsar
age.  Measurements  of  their  proper  motion  and  the  detection  of
pulsations with the pulsar periods in the optical range 
would be most important to provide
firm  evidence   of  the  pulsar   nature  of  the   detected  optical
objects. Simultaneous  studies of the optical and  X-ray pulse profile
would  provide  stronger indications  whether  the  optical and  X-ray
emissions 
are generated by
the same physical process.  Our observations of a
very faint  \psr \ with the  Subaru show that new  generation of large
ground-based telescopes is very  effective for these studies and could
lead to a  considerable increase of the number  of pulsars detected in
the optical range in the near future.

\acknowledgements This work was  supported in part by CONACYT projects
25454-E and 36585-E, RFBR  grants 02-02-17668 and 00-07-90183.  We are
grateful  to Y.~Komiyama  for the  help during  observations  with the
Subaru,  and  to G.~Pavlov  for  the  unpublished  results on  revised
astrometrical referencing  of the \hst/FOC  image of the  \psr\ field.
Some  of the  data  presented in  this  paper were  obtained from  the
Multimission Archive at the  Space Telescope Science Institute (MAST).
STScI is operated  by the Association of Universities  for Research in
Astronomy, Inc., under NASA contract NAS5-26555.  Support for MAST for
non-HST data is provided by the NASA Office of Space Science via grant
NAG5-7584 and by other grants  and contracts.  
ABK is grateful to the Astronomy Department of the University of
Washington for hospitality.
We are also grateful to M.~Richer for carefull reading of this text 
and useful remarks,
and to V.~Palshin for a discussion.


\begin{thebibliography}{}

\bibitem[Anderson et al., (1993)]{Anderson} Anderson, S.B., C\'{o}rdova, F.A., Pavlov, G.G.,
 Robinson, C.R., Tompson, R. J., 1993, ApJ, 414, 867 

\bibitem[Arons 1981]{Arons} Arons, J. 1981, ApJ, 248, 1099

\bibitem[Becker \& Tr\"{u}mper 1997]{Becker}Becker W.,  Tr\"{u}mper J., 1997
A\&A, 326, 682

\bibitem[Bell Burnell 1998]{Bell} Bell Burnell, S.J., 1998, in:  Neutron Stars and Pulsars,  
eds. N. Shibazaki, N. Kawai, S. Shibata, T. Kifune, Univ. Acad. Press Inc., Tokyo, p1   

\bibitem[Bignami et al., 1993]{Bignami} Bignami, G.F., Caraveo, P.A., Mereghetti, S., 1993
Nature, 361, 704

\bibitem[Brisken et al. (2000)]{Brisken0}Brisken, W F., Benson, J. M.,
 Beasley, A. J., Fomalont, E, B., Goss, W. M., 
Thorsett, S. E.,  2000, ApJ, 541, 959 

\bibitem[2002]{Brisken2}
Brisken, W.F., Benson, J.M., Goss, W.M., Thorsett, S.E.
2002, ApJ, 571, 906


\bibitem[Caraveo et al., (1996)]{Caraveo2} 
Caraveo, P. A., Bignami, G. F., Mignani, R., Taff, L. G., 1996, ApJ, 461, L91

\bibitem[Caraveo et al.,  (2001)]{Caraveo}  Caraveo, P.A.,  De Luca, A., 
Mignani, R.P.,  Bignami, C.F., 2001,  ApJ, 561, 930

\bibitem[Cordova et al. (1989)]{Cordova} 
Cordova, F.A., Hjellming, R.M., Mason, K.O., \& Middleditch, J.  1989 ApJ, 345, 451 

\bibitem[2001]{Crusius}Crusius-W\"atzel, A R., Kunzl, T.,
 Lesch, H., 2001, ApJ,  546, 401

\bibitem[Finley et al., 1992]{Finley} Finley. J.P., \"{O}gelman, H., Kiziloglu, \"{U}, 1992, ApJ.,
394, L21

\bibitem[Fomalont et al.~(1992)]{Fomalont}Fomalont, E. B., Goss, W. M.,
 Lyne, A. G., Manchester, R. N., Justtanont, K., 1992, MNRAS, 258, 497


\bibitem[Gwinn et al.~(1986)]{Gwinn}Gwinn, C. R., Taylor, J. H.,
 Weisberg, J. M.,Rawley, L. A., 1986, AJ, 91, 338

\bibitem[Hill et al.~(1997)]{Hill}  
 Hill, Robert J.; Dolan, Joseph F.; Bless, Robert C.; Boyd, Patricia T.;
                     Percival, Jeffrey W.; Taylor, M. J.; van Citters, G. W., 1997, ApJ, 486, L99 



\bibitem[2001]{Koptsevich}Koptsevich A., B., Pavlov G. G., 
Zharikov S. V., Sokolov V. V.,  Shibanov Yu. A., Kurt V. G., 2001, A\&A, 370, 1004 

\bibitem[Kurt et al.~1998]{Kurt}Kurt, V. G., Sokolov, V. V., Zharikov, S. V.,
Pavlov, G. G., Komberg, B. V., 1998, A\&A, 333, 547

\bibitem[Kurt et al.~(2000)]{Kurt1} Kurt V.G,  Komarova V.N,  Fatkhullin T.A., Sokolov V.V,  
Koptsevich A. B.,  Shibanov Yu. A., 2000, Bulletin SAO RAS, 49, 5

\bibitem[Landolt~1992]{Landolt}Landolt A.,  1992, AJ, 104, 340

\bibitem[1994]{Manning}Manning, R. A., Willmore, A. P.,
 1994, MNRAS, 266, 635

\bibitem[Middleditch et al., 1987]{Middleditch} Middleditch, J.;
 Pennypacker, C. R.; Burns, M. S.,  1987, ApJ, 315,  142

\bibitem[Mignani et al. 2000]{Mignani2} Mignani, R. P.; Caraveo, P. A.; Bignami, G. F., 
 2000, Msngr, 99, 22

\bibitem[2001]{Mignani3}  Mignani, R. P., Caraveo, P. A. 2001, A\&A, 376, 213

\bibitem[Miyazaki et al. 1998]{Miyazaki}
Miyazaki, S., Sekiguchi, M. Imi, K., Okada, N.,
Nakata, F., Komiyama, Y., 1998, SPIE, 3355, 363  

\bibitem[Miralles et al. 1998]{Miralles} Miralles, J. A., Urpin, V., Konenkov, D., 1998, 
ApJ, .503, 368

\bibitem[Nasuti et al., 1997]{Nasuti}Nasuti, F. P., Mignani, R.,
Caraveo, P. A.,
Bignami, G. F.,  1997, A\&A, 323, 839

\bibitem[1996]{Pavlov} Pavlov, G. G., Stringfellow, G. S., Cordova, F. A., 1996, ApJ, 467, 370

\bibitem[2000]{Pavlov1} Pavlov G.G., 2000, private communication

\bibitem[Percival et al.~(1993)]{Percival}Percival, J. W., Biggs, J. D.,
 Dolan, J. F., Robinson, E. L.,
 Taylor, M. J., Bless, R. C.,
 Elliot, J. L., Nelson, M. J.,
 Ramseyer, T. F.,
 van Citters, G. W., Zhang, E., 1993, ApJ, 407, 276

\bibitem[Seward \& Wang (1988)]{Seward}Seward F., \& Wang Z., 1988, Ap.J.,
 332, 199

\bibitem[Shibata 1995]{Shibata} Shibata S., 1995, MNRAS, 276, 537 

\bibitem[Sokolov et al.~(1998)]{Sokolov}Sokolov, V. V., Kurt, V. G., Zharikov, S. V.,
 Shibanov, Yu. A., Koptsevich, A. B., 1998, 
The 19th Texas Symposium on Relativistic Astrophysics and Cosmology, held in Paris,
 France, Dec. 14-18, 1998. Eds.: J. Paul, T. Montmerle, and E. Aubourg (CEA Saclay).

\bibitem[Taylor et al.\ 1993]{Taylor}
Taylor, J.H., Manchester, R.N., \& Lyne, A.G.,   
1993, ApJS, 88, 529.

\bibitem[Thompson, 2000]{Thompson} Thompson D., 2000, Advances in Space Research, 25, 659


\bibitem[1997]{Wang}Wang, F. Y.-H., 
Halpern, J. P., 1997, ApJ,482, L159


\bibitem[2000]{We} 
Weisskopf, M. C., Hester, J.J,  Tennant, A.F., Elsner, R.F.,  Schulz, N. S.,  Marshall, H.L., 
Karovska, M.,  Nichols, J.S.,  Swartz, D.A., 
Kolodziejczak, J.J.,  O'Dell, S.L., 2000, ApJ, 536,  L81

\bibitem[Yakovlev et al. 1999]{Yakovlev} Yakovlev, D.G., Levenfish, K.P., 
 Shibanov Yu.A.,  1999, Physics-Uspekhi 42, 737 

\bibitem[1996]{Zavlin} Zavlin V. E., Pavlov G. G., Shibanov Yu. A., 1996, A\&A,  315, 141 


\end{thebibliography}
\end{document}